\documentclass[aps,prb,twocolumn]{revtex4}
\usepackage{amsmath}
\usepackage{color}
\usepackage{amssymb}

\usepackage{graphicx}

\begin{document}

\title{ Low-temperature resistance  in metals without inversion center }

\author{V.P.Mineev$^{1,2}$}
\affiliation{$^1$Univ. Grenoble Alpes, CEA, INAC, PHELIQS, GT, F-38000 Grenoble, France\\
$^2$Landau Institute for Theoretical Physics, 142432 Chernogolovka, Russia}

\begin{abstract}
The well known quadratic low-temperature dependence of resistance in ordinary metals habitually serves as the criterium of applicability of the Landau Fermi liquid theory to the description of  electron liquid in  concrete material.
Such a type of behavior is determined by momentum relaxation due to the electron-electron scattering.
Here I consider 
this problem  in the metals without inversion center. It is shown that the corresponding scattering time 
at temperatures much smaller than the spin-orbit coupling is  practically temperature independent. 
\end{abstract}

\date{\today}
\maketitle

\section{Introduction}

Normal metal resistivity at low temperatures is usually described by the well known formula
\begin{equation}
\rho=\rho_0+AT^2.
\label{LP}
\end{equation}
This dependence derived by Landau and Pomeranchuk \cite{Landau1936} is determined by the relaxation time
due to electron-electron scattering 
\begin{equation}
\tau_{ee}\propto \frac{1}{T^2},
\label{tau}
\end{equation}
which is faster than the case of electron-phonon interaction at temperatures  much lower than the Debye temperature 
\begin{equation}
\tau_{eph}\propto \frac{1}{T^5}.
\end{equation}
The Landau-Pomeranchuk's relaxation is caused by the interaction between electrons. They, owing to the Pauli principle, can scatter with each other 
only in a narrow energy layer of the order of temperature near  the Fermi surface.
The realization  of dependence (\ref{LP}) in a concrete material  is commonly used as the direct indication of applicability
of the Landau Fermi-liquid theory, and vice versa, a deviation from this dependence stimulates the search of some physical mechanism responsible
for non-Fermi-liquid behavior.

In neutral  Fermi liquid like normal $^3$He the low-temperature relaxation due to quasiparticle-quasiparticle  scattering has the same $1/T^2$ dependence that determines the temperature dependence of viscosity, thermal conductivity \cite{1,2} and the longitudinal spin-diffusion
coefficient \cite {3}. The latter takes place even in $^3$He polarized by an external magnetic field $H$ where the Fermi spheres of quasiparticles with opposite spins have different radii.
Nevertheless, during and after the scattering processes  spin-up and spin-down quasiparticles remain at a distance of the order of temperature $T$ from  their Fermi surfaces. 

Quite a different situation is realised for the  spin diffusion in the direction perpendicular to an external magnetic  field  \cite{Mullin1988,Meyerovich1990}. Here, the  scattering processes involve all the states between two Fermi surfaces, and the relaxation time acquires the following temperature dependence
\begin{equation}
{\tau}_\perp\propto\frac{1}
{(2\pi T)^{2}+(g H)^{2}},
\label{e38}
\end{equation}
where $g$ is  the $^3$He nuclei gyromagnetic ratio. 
This result has been obtained \cite{Mineev2004}  making  use  of the kinetic equation for the matrix distribution function
of Fermi particles  \cite{Silin1957,Silin1971,Mullin1988})  in the frame of Landau Fermi-liquid theory  
 and does not 
originate  no matter  which  violation of it.

Recently there was   interest in substances without the center of inversion \cite{Bauer}.  In this case   the spin-orbit coupling lifts the spin degeneracy of the electron energy bands. The initial Fermi surface splits on two Fermi surfaces for the electrons characterized by the opposite helicity that is the spin projection on the direction of the momentum dependent vector of spin-orbit coupling. The difference of the Fermi momenta of these Fermi surfaces $\Delta k_F$ is determined by the value of spin-orbital coupling.
Like the transverse spin relaxation in the spin-polarized $^3$He one can expect that in noncentrosymmetric metals the relaxation time due to  electro-electron
interaction have the following form
\begin{equation}
{\tau}_{ee}\propto\frac{1}
{(2\pi T)^{2}+(v_F\Delta k_F)^{2}}.
\label{39}
\end{equation}

Here we derive the formula for low-temperature conductivity in metals without inversion center determined by the momentum relaxation due to the electron-electron scattering. It has  more complex structure than that
given by the simple relation $\sigma\propto \tau_{ee}$. However,  it is also determined  
  by the combination similar to the denominator of 
Eq.(\ref{39}).
The spin-orbital band splitting $v_F\Delta k_F$ is directly expressed through   the corresponding splitting of de Haas - van Alphen magnetization oscillation frequencies \cite{Mineev2005}.
Determined experimentally the typical magnitude of band splitting  in many noncentrosymmetric metals is of the order of hundreds Kelvin \cite{Terashima2008, Onuki2014,Maurya2018}. This is much less than the Fermi energy but comparable with the Debye temperature. Hence   at temperatures significantly lower than the Debye temperature the $T^2$ term in the denominator of Eq.({\ref{39}) is proved to be unobservable. The calculations in the paper are performed under assumption $v_F\Delta k_F\ll\varepsilon_F$.

 The paper is organized as follows.  Section II contains the basic notions of the electron energy spectrum and the equilibrium distribution in metals without inversion. In Section III the formula for the low-temperature conductivity in noncentrosymmetric metals is derived. 
The solution of the kinetic equation used in this derivation is obtained in the Appendix at the end of the paper.
The Appendix is divided into  two parts: the first contains the kinetic equation and its general properties, the second one is devoted to the derivation of deviation of distribution function from equilibrium under an electric field.
 The conclusion and discussion of obtained results are given in Section IV.

\section{Equilibrium distribution of electrons}

 The spectrum of noninteracting
electrons in a metal without inversion center is:
\begin{equation}
\label{H_0}
 \hat \varepsilon({\bf k})-\mu\hat 1
 = \xi({\bf k})\hat 1+\mbox{\boldmath$\gamma$}({\bf k}) 
   \cdot \mbox{\boldmath$\sigma$}
\end{equation}
where 
$\xi({\bf k})=\varepsilon({\bf k})-\mu$ denotes the spin-independent part of the spectrum measured relative to the chemical potential $ \mu$,
$\hat 1$ is the unit $2\times 2$ matrix in the spin space, $\mbox{\boldmath$\sigma$}=(\sigma_x,\sigma_y,\sigma_z)$ are the Pauli matrices. 
The second term in Eq.
(\ref{H_0}) describes the  spin-orbit  coupling whose form depends on the specific noncentrosymmetric crystal structure .
The pseudovector $\mbox{\boldmath$\gamma$}({\bf k})$  satisfies
$\mbox{\boldmath$\gamma$}(-{\bf k})=-\mbox{\boldmath$\gamma$}({\bf k})$ and 
$g\mbox{\boldmath$\gamma$}(g^{-1} {\bf k})=\mbox{\boldmath$\gamma$}({\bf k})$,
where $g$ is any symmetry operation in the  point group ${\cal G}$ of
the crystal. A more detailed  theoretical description of noncentrosymmetric metals is presented in the paper \cite{Mineev2012}. 
We shall work  with isotropic spectrum $\varepsilon({\bf k})=\frac{k^2}{2m}$ and 
\begin{equation}
\label{gamma_O}
    \mbox{\boldmath$\gamma$}({\bf k})=\gamma{\bf k},
\end{equation}
compatible with the 3D cubic crystal symmetry. Here $\gamma$ is a constant. The calculations with anisotroppic spectrum and  $ \mbox{\boldmath$\gamma$}({\bf k})$
corresponding to other crystal symmetries are much more cumbersome but do not qualitatively change the results.

The eigenvalues of the matrix (\ref{H_0}) are
\begin{equation}
    \xi_{\pm}({\bf k})=\xi({\bf k})\pm
    |\mbox{\boldmath$\gamma$}({\bf k})|.
\label{e3}
\end{equation}
There are two Fermi surfaces determined by the equations
\begin{equation}
\label{e4}
    \xi_{\pm}({\bf k})=0
\end{equation}
with different Fermi momenta
\begin{equation}
k_{F\pm}=\mp m \gamma+\sqrt{2m\mu+(m\gamma)^2}
\end{equation}
and common value of the Fermi velocity 
\begin{equation}
{\bf v}_{F}=\frac{\partial(\varepsilon \pm\gamma k)}{\partial {\bf k}}|_{k=k_{F\pm}}=\hat{\bf k}\sqrt{\frac{2\mu}{m}+\gamma^2},
\end{equation}
here $\hat{\bf k}$ is the unit vector along momentum ${\bf k}$.

The matrix of equilibrium electron distribution function is
\begin{equation}
\hat n_0
=\frac{n_++n_-}{2}\hat 1+\frac{n_+-n_-}{2|\mbox{\boldmath$\gamma$}|} \mbox{\boldmath$\gamma$} \cdot \mbox{\boldmath$\sigma$},
\label{eqv}
\end{equation}
where 
\begin{equation}
n_\pm=\frac{1}{e^{\xi_{\pm}}+1}
\end{equation}
are the Fermi functions.
Near the corresponding Fermi surfaces the dispersion laws are
\begin{equation}
\xi_\pm\approx v_F(k-k_{F\pm})=\varepsilon-\mu_\pm,
\end{equation}
with 
\begin{equation}
\varepsilon=v_Fk,~~~~~\mu_\pm=v_Fk_{F\pm},~~~~~~\mu_+-\mu_-=-2mv_F\gamma.
\label{ene}
\end{equation}

\section{Conductivity}

The matrix of non-equlibrium   distribution function is  the sum of the scalar and the spinor parts
\begin{equation}
\hat n=\frac{1}{2}\left[ f~ \hat 1+{\bf g}\cdot\mbox{\boldmath$\sigma$}  \right ].
\end{equation}

The current density is
\begin{widetext}
\begin{eqnarray}
{\bf j}=e~Tr\int\frac{d^3{\bf k}_1}{(2\pi)^3}\frac{\partial  \hat \varepsilon({\bf k}_1)}{\partial{\bf k}_1}\hat n=
\frac{1}{2}e~Tr\int\frac{d^3{\bf k}_1}{(2\pi)^3}\left (\frac{\partial \xi({\bf k}_1)}{\partial {\bf k}_1}+\frac{\partial \mbox{\boldmath$\gamma$}({\bf k}_1)\cdot
\mbox{\boldmath$\sigma$}}{\partial {\bf k}_1}\right )\left (\delta  f({\bf k}_1) \hat 1+
\delta {\bf g}({\bf k}_1)\cdot\mbox{\boldmath$\sigma$}
  \right )\nonumber\\
 =
 e\int\frac{d^3{\bf k}_1}{(2\pi)^3}\left (\frac{{\bf k}_1}{m}\delta  f({\bf k}_1) +\gamma
 \delta{\bf  g}({\bf k}_1)                \right ).~~~~~~~~~~~~~~~~~~~~~~~~~~~~~~~~~~~~
 \label{j}
\end{eqnarray}
\end{widetext}
Here, 
\begin{eqnarray}
\delta f=f-(n_++n_-),\\
\delta{\bf g}={\bf g}-\frac {\mbox{\boldmath$\gamma$}}{ |\mbox{\boldmath$\gamma$}|}(n_+-n_-)
\end{eqnarray} 
are the deviations of scalar and spinor distribution functions from the equilibrium distribution function.
The derivation of them from the kinetic equation  performed under assumption $v_F\Delta k_F\ll\varepsilon_F$
 is presented in the Appendix B. They are
 \begin{widetext}
 \begin{equation}
\delta f({\bf k}_1)=-\frac{4e}{\pi \tilde W_0 m^3
I_1(\varepsilon_1)} {\bf E}\cdot{\bf v}_1
\frac{\partial(n_{1+}+n_{1-})}{\partial \varepsilon_1},
\label{scalar1}
\end{equation}
\begin{equation}
\delta{\bf g}({\bf k}_1)=-\frac{4e}{\pi m^3\left [\tilde W_0
I_1(\varepsilon_1)-W_0I_2\right ]}\left [{\bf E}\cdot{\bf v}_1\frac{\partial(n_{1+}-n_{1-})}{\partial \varepsilon_1}\hat{\bf k}_1+
\frac{(n_{1+}-n_{1-)}}{|{\bf k}_1|}\left  ({\bf E}-({\bf E}\cdot\hat{\bf k}_1)\hat{\bf k}_1  \right )\right ],
\label{spinor1}
\end{equation}
where the integrals $I_1(\varepsilon_1),I_2$  are given by Eqs. (\ref{I1}) and (\ref{I2}).
Substituting these expressions in Eq.(\ref{j}) we obtain
 \begin{eqnarray}
{\bf j}=\frac{4e^2}{3\pi m^3
}\left \{v_F\frac{N_{0+}k_{F+}+N_{0-}k_{F-}}{m\tilde W_0I_1(\mu_+)}\right.~~~~~~~~~~~~~~~~~~~~~~~~~~~~~~~\nonumber\\
\left.+\gamma\left [ \frac{N_{0+}v_F}{\left [\tilde W_0I_1(\mu_+)-W_0I_2\right ]}- \frac{N_{0-}v_F}{\left [\tilde W_0I_1(\mu_-)-W_0I_2\right]}-2\int
\frac{d^3{\bf k}_1}{(2\pi)^3}
\frac{n_{1+}-n_{1-}}{k_1\left[\tilde W_0I_1(\varepsilon_1)-W_0I_2\right ]}
\right]\right \}{\bf E},
\end{eqnarray}
where 
$
N_{0\pm}=\sqrt{2m^3\mu_\pm}/2\pi^2
$
are the density of states at the corresponding Fermi momenta.
Thus, the conductivity is 
 \begin{eqnarray}
\sigma=\frac{4e^2}{3\pi m^3
}\left \{v_F\frac{N_{0+}k_{F+}+N_{0-}k_{F-}}{m\tilde W_0I_1(\mu_+)}\right.~~~~~~~~~~~~~~~~~~~~~~~~~~~~~~~~~~~~~~\nonumber\\
\left.+\gamma\left [ \frac{N_{0+}v_F}{\left [\tilde W_0I_1(\mu_+)-W_0I_2\right]}- \frac{N_{0-}v_F}{\left [\tilde W_0I_1(\mu_-)-W_0I_2\right ]}-2\int
\frac{d^3{\bf k}_1}{(2\pi)^3}
\frac{n_{1+}-n_{1-}}{k_1\left [\tilde W_0I_1(\varepsilon_1)-W_0I_2\right ]}
\right]\right \}.
\label{sigma}
\end{eqnarray}
\end{widetext}
It consists of  two parts:  the first line corresponds to the linear in field deviation of the scalar part of the distribution function whereas the second line originates from the  deviation of  the spinor part of the distribution function.

In the absence of spin-orbit interaction $\gamma=0$
\begin{equation}
I_1(\mu_+=\mu_-)=2(\pi T)^2, ~~~~~~I_2=0
\end{equation}
 we come to the Landau-Pomeranchuk temperature dependence of conductivity
\begin{equation}
\sigma=\frac{4e^2v_F^2N_0}{3\pi^3m^3\tilde W_0}\frac{1}{T^2}.
\label{LanPom}
\end{equation}
On the other hand at finite Fermi radii splitting \\$|\mu_+-\mu_-|>>T$ the  integrals
\begin{equation}
I_1(\mu_+)=I_1(\mu_-)\propto I_2\propto (\mu_+-\mu_-)^2
\end{equation}
are practically temperature independent  and the conductivity is determined by the temperature independent 
Eq.(\ref{sigma}).

\section{Concluding remarks}

We have derived  the low-temperature conductivity in the metals without inversion center determined by the 
momentum relaxation due to electron-electron scattering. Unlike to the center-symmetric metals the conductivity in the materials with the space parity violation  proved to be temperature independent so long as $T\ll v_F\Delta k_F\ll\varepsilon_F$.  The physical reason for this behavior is that in noncentrosymmetric metals the processes of relaxation are determined by all the electronic states in between the Fermi surfaces split by the spin-orbit coupling. 

At $\gamma=0$ the Landau-Pomeranchuk electron-electron scattering rate determining the conductivity Eq. ({\ref{LanPom}) is
\begin{equation}
\frac{1}{\tau_{ee}}\sim\frac {V^2}{\varepsilon_F ^2}\frac {T^2}{\varepsilon_F},
\end{equation}
where $V$ is the amplitude of screened short range potential of the electron-electron interaction.  At low temperatures, when the quasiparticles energy $\xi$ counted from the Fermi energy is of the order of temperature, the inequality  
$1/\tau_{ee}\propto \xi^2/\varepsilon_F\ll\xi$ serves as the base of the Landau Fermi liquid theory.

The corresponding scattering rate at finite Fermi radii splitting $|\mu_+-\mu_-|>>T$ determining the conductivity Eq. ({\ref{sigma}) is
\begin{equation}
\frac{1}{\tau_{ee}}\sim\frac {V^2}{\varepsilon_F ^2}\frac {(\mu_+-\mu_-)^2}{\varepsilon_F}.
\end{equation}
So,  the condition $1/\tau_{ee}\propto (\mu_+-\mu_-)^2/\varepsilon_F\ll\xi$ is fulfilled for not arbitrary small $\xi$ values. Thus, the problem of  a life-time of quasiparticles  interacting by the short range potential in metals without inversion center demands  further investigation.

The presented calculations made for the  noncentrosymmetric  metals with cubic symmetry. In this case the spin degeneracy is completely lifted, and the low- temperature dependence of resistivity due to electron-electron scattering is 
\begin{equation}
\rho_{ee}-\rho_{ee}(T=0)\propto T^2.
\end{equation}
In less symmetric materials  the degeneracy of states is also lifted everywhere
besides isolated points where the two split Fermi surfaces touch each other as this is the case in the metals with tetragonal or tetrahedral symmetry (see Ref.15). This residual degeneracy  can lead to the negligibly small temperature dependent correction to 
 the $T^2$ temperature dependence  of  resistivity originating from the electron-electron scattering. 

In a metal without inversion center the total resistivity at zero temperature consists of two 
 parts originating  from resistivity due to electron-electron scattering  and 
due to electron scattering on impurities
\begin{equation}
\rho=\rho_{ee}(T=0)+\rho_{imp}.
\end{equation}
The resistivity due impurity scattering is proportional to  impurity concentration $\rho_{imp}\propto n_{imp}$.
Thus, the zero-temperature resistivity due to electron-electron scattering $\rho_{ee}(T=0)$ can be experimentally found by the measuring of low temperature resistivity $\rho (n_{imp})$ at several finite impurity concentrations with subsequent taking the formal limit
\begin{equation}
\rho_{ee}(T=0)=\rho (n_{imp} \to 0).
\end{equation}

\appendix
\section{Kinetic equation}

The  kinetic equation for  the matrix  distribution function of electrons derived by V.P.Silin \cite{Silin1957} is
\begin{eqnarray}
\frac{\partial \hat n_1}{\partial t}-i[\hat\varepsilon_1,\hat n_1]+\frac{1}{2}\left ( \frac{\partial\hat \varepsilon_1}{\partial {\bf k}_1}
\frac{\partial\hat n_1}{\partial {\bf r}} +\frac{\partial\hat n_1}{\partial {\bf r}} \frac{\partial\hat \varepsilon_1}{\partial {\bf k_1}}  \right )\nonumber\\-
\frac{1}{2}\left ( \frac{\partial\hat \varepsilon_1}{\partial {\bf r}}
\frac{\partial\hat n_1}{\partial {\bf k}_1} +\frac{\partial\hat n_1}{\partial {\bf k}_1} \frac{\partial\hat \varepsilon_1}{\partial {\bf r}}  \right )=\hat I,~~~~~~~~~~
\end{eqnarray}
where $[\hat\varepsilon_1,\hat n_1]$ is the commutator of $\hat\varepsilon_1=\hat\varepsilon({\bf k}_1)$ and $\hat n_1=\hat n({\bf k}_1)$. We put $\hbar=1$. The  integral term for particle-particle  collisions in the Born approximation was derived  by Silin ( Ref.9) and Jeon and Mullin (Ref.5).
Including the Umklapp processes of scattering it is 
\begin{widetext}
\begin{equation}
\hat I=2\pi\int d^3{\bf k}^\prime\frac{d^3{\bf k}^{\prime\prime}}{(2\pi)^3}\frac{d^3{\bf k}_2}{(2\pi)^3}\sum_{\bf m}
\delta(\varepsilon_{1}+\varepsilon_{2}-\varepsilon^\prime-
\varepsilon^{\prime\prime}) \delta\left({\bf k}_1+{\bf k}_{2}-{\bf k}^\prime-{\bf
k}^{\prime\prime}-\frac{2\pi{\bf m}}{a}\right) \hat F,
\label{e27}
\end{equation}
where $\frac{2\pi{\bf m}}{a}$ is a vector of reciprocal lattice,
\begin{eqnarray}
\hat F
 =
\frac{1}{2}W_1\left \{ [ \hat n^\prime,(\hat 1-\hat n_1) ]_+Tr((\hat 1-\hat n_2)n^{\prime\prime})-
 [(\hat 1- \hat n^\prime),\hat n_1 ]_+Tr (\hat n_2(\hat 1-\hat n^{\prime\prime}))\right \}\nonumber\\
+\frac{1}{2}
W_2\left\{[\hat n^\prime(\hat 1-\hat n_2)\hat n^{\prime\prime},(\hat 1-\hat n_1) ]_+-  [(\hat 1-\hat n^\prime)\hat n_2(\hat 1-\hat n^{\prime\prime}),\hat n_1 ]_+  \right \}.
\label{e28}
\end{eqnarray}
\end{widetext}
Here $[\hat A,\hat B]_+$ means the anticommutator, and the following designations $\hat n^\prime=\hat n({\bf k}^\prime),~ \varepsilon^\prime=\varepsilon ({\bf k}^\prime) $ etc are introduced.  In the isotropic Fermi liquid like $^3$He $W_1=[V(|{\bf k}_1-{\bf k}^\prime|)]^2,~~~~
W_2=-V(|{\bf k}_1-{\bf k}^\prime|)V(|{\bf k}_1-{\bf k}^{\prime\prime}|)$ are expressed trough the Fourier transform of the quasiparticles potential of interaction. The latter in concrete metal is unknown and due to  charge screening we put them as the constants: $W_1=W_0/2,~~W_2=-W_0/2$.

The matrix of equilibrium electron distribution function is
\begin{equation}
\hat n_0=\hat P_+n_++\hat P_-n_-
\label{equa}
\end{equation}
where 
\begin{equation}
\hat P_\pm({\bf k})=\frac{1}{2}\pm
\frac{\mbox{\boldmath$\gamma$({\bf k})}
\cdot\mbox{\boldmath$\sigma$}}{2|\mbox{\boldmath$\gamma$}({\bf k})|}
\end{equation}
are the projection operators such that $$\hat P_\pm^2=\hat P_\pm,~~\hat P_+\hat P_-=\hat P_-\hat P_+=0,~~\hat P_++\hat P_-=1.$$
The substitution of equilibrium distribution Eq.(\ref{equa})  causes the collision integral  to vanish. Indeed, substituting Eq.(\ref{equa}) to Eq. (\ref{e28}) we obtain
\begin{widetext}
\begin{eqnarray}
\hat F=\frac{1}{4}W_0\sum_{\lambda,\mu,\nu,\tau}\left \{\left [n_\lambda^\prime (1-n_{1\mu})(1-n_{2\nu})n_\tau^{\prime\prime}-
(1-n_\lambda^\prime)n_{1\mu}n_{2\nu}(1-n_{2\tau}^{\prime\prime}) \right ] \left (\hat P_\lambda^\prime\hat P_{1\mu}+\hat P_{1\mu}\hat P_\lambda^\prime\right )Tr\left (\hat P_{2\nu}\hat P_{\tau}^{\prime\prime} \right )\right.\nonumber\\
-\left. [n_\lambda^\prime(1-n_{2\mu})n^{\prime\prime}_\nu(1-n_{1\tau})-(1-n_\lambda^\prime)n_{2\mu}(1-n^{\prime\prime}_\nu)n_{1\tau}] 
\left ( \hat P_\lambda^\prime \hat P_{2\mu} \hat P_{\nu}^{\prime\prime} \hat P_{1\tau}+ \hat P_{1\tau}\hat P_\lambda^\prime \hat P_{2\mu} \hat P_{\nu}^{\prime\prime}
 \right )\right\}
\end{eqnarray}
\end{widetext}
 The substitution of this expression to the Eq.(\ref{e27})  yields zero because the combination in square parenthesis 
 $\left [n_\lambda^\prime (1-n_{1\mu})(1-n_{2\nu})n_\tau^{\prime\prime}-
(1-n_\lambda^\prime)n_{1\mu}n_{2\nu}(1-n_{2\tau}^{\prime\prime}) \right ]$ is equal to zero at arbitrary $\lambda, \mu,\nu,\tau=\pm$ due to the energy conservation $\varepsilon_{1}+\varepsilon_{2}=\varepsilon^{\prime}+
\varepsilon^{\prime\prime}$.

The conservation laws of total particles number
$\int\frac{d^3{\bf k}_1}{(2\pi)^3}Tr(\hat n_1) $, spin $\int\frac{d^3{\bf k}_1}{(2\pi)^3}Tr(\mbox{\boldmath$\sigma$}\hat n_1) $, momentum $\int\frac{d^3{\bf k}_1}{(2\pi)^3}Tr({\bf k}_1\hat n_1) $ and energy $\int\frac{d^3{\bf k}_1}{(2\pi)^3}Tr(\hat \varepsilon({\bf k}_1) \hat n_1) $ are satisfied.
The check of  their validity  using the Eqs. (A1), (A2) and (A3)  belongs to that category of calculations which are more easily done independently than by following their development.

 \section{Deviations of  distribution functions from the equilibrium distribution}
 
In a constant electric field the stationary  kinetic equation acquires the form
\begin{equation}
\left (e{\bf E}\cdot\frac{\partial}{\partial{\bf k}_1}\right)\hat n_1
=\hat I
\label{k}
\end{equation}
The matrix Fermi distribution function is convenient to represent as the sum of the scalar and the spinor parts
\begin{equation}
\hat n=\frac{1}{2}\left[ f~ \hat 1+{\bf g}\cdot\mbox{\boldmath$\sigma$}  \right ].
\end{equation}
The kinetic equations equations for this function are
\begin{equation}
\left (e{\bf E}\cdot\frac{\partial}{\partial{\bf k}_1}\right)f_1
=Tr\hat I,
\label{Tr}
\end{equation}
\begin{equation}
\left (e{\bf E}\cdot\frac{\partial}{\partial{\bf k}_1}\right){\bf g}_{1}=Tr(\mbox{\boldmath$\sigma$}\hat I),
\label{Trs}
\end{equation}
where (see Ref.5)
\begin{widetext}
\begin{eqnarray}
Tr\hat I=\frac{\pi}{2}W_0\int d^3{\bf k}^\prime\frac{d^3{\bf k}^{\prime\prime}}{(2\pi)^3}\frac{d^3{\bf k}_2}{(2\pi)^3}\sum_{\bf m}
\delta(\varepsilon_{1}+\varepsilon_{2}-\varepsilon^\prime-
\varepsilon^{\prime\prime}) \delta\left({\bf k}_1+{\bf k}_{2}-{\bf k}^\prime-{\bf
k}^{\prime\prime}-\frac{2\pi{\bf m}}{a}\right)\times~~~~~~~~~~~~~~~~~~\nonumber\\
\left \{\left [f^\prime-\frac{1}{2}(f_1f^\prime+{\bf g}_1\cdot{\bf g}^\prime)\right ] 
\left [f^{\prime\prime}-\frac{1}{2}(f_2f^{\prime\prime}+{\bf g}_2\cdot{\bf g}^{\prime\prime})\right ]-\left [f_1
-\frac{1}{2}(f_1f^\prime+{\bf g}_1\cdot{\bf g}^\prime)\right ]
\left [f_2-\frac{1}{2}(f_2f^{\prime\prime}+{\bf g}_2\cdot{\bf g}^{\prime\prime})\right ]   \right.\nonumber\\
\left.-\left [{\bf g}^\prime-\frac{1}{2}(f^\prime{\bf g}_1+f_1{\bf g}^\prime)\right ]\cdot 
\left [{\bf g}^{\prime\prime}-\frac{1}{2}(f^{\prime\prime}{\bf g}_2+f_2{\bf g}^{\prime\prime})\right ]
+\left [{\bf g}_1-\frac{1}{2}(f^\prime{\bf g}_1+f_1{\bf g}^\prime)\right ]\cdot
\left [{\bf g}_2-\frac{1}{2}(f^{\prime\prime}{\bf g}_2+f_2{\bf g}^{\prime\prime})\right ]  \right \},~~~~
\label{30}
\end{eqnarray}
\begin{eqnarray}
Tr(\mbox{\boldmath$\sigma$}\hat I)=\frac{\pi}{2}W_0\int d^3{\bf k}^\prime\frac{d^3{\bf k}^{\prime\prime}}{(2\pi)^3}\frac{d^3{\bf k}_2}{(2\pi)^3}\sum_{\bf m}
\delta(\varepsilon_{1}+\varepsilon_{2}-\varepsilon^\prime-
\varepsilon^{\prime\prime}) \delta\left({\bf k}_1+{\bf k}_{2}-{\bf k}^\prime-{\bf
k}^{\prime\prime}-\frac{2\pi{\bf m}}{a}\right)\times~~~~~~~~~~~~~~~~~~\nonumber\\
\left \{\left [f^{\prime\prime}-\frac{1}{2}(f_2f^{\prime\prime}+{\bf g}_2\cdot{\bf g}^{\prime\prime})\right ] 
\left [{\bf g}^{\prime}-\frac{1}{2}(f^\prime{\bf g}_1+f_1{\bf g}^{\prime})\right ]-
\left [f_2-\frac{1}{2}(f_2f^{\prime\prime}+{\bf g}_2\cdot{\bf g}^{\prime\prime})\right ]
\left [{\bf g}_1-\frac{1}{2}(f^\prime{\bf g}_1+f_1\cdot{\bf g}^{\prime})\right ]  \right.\nonumber\\
\left.-\left [f^\prime-\frac{1}{2}(f_1f^\prime+{\bf g}_1\cdot{\bf g}^\prime)\right ] 
\left [{\bf g}^{\prime\prime}-\frac{1}{2}(f^{\prime\prime}{\bf g}_2+f_2{\bf g}^{\prime\prime})\right ]
+\left [f_1-\frac{1}{2}(f_1f^\prime+{\bf g}_1\cdot{\bf g}^\prime)\right ]
\left [{\bf g}_2-\frac{1}{2}(f^{\prime\prime}{\bf g}_2+f_2{\bf g}^{\prime\prime})\right ]  \right.~~~~\nonumber\\
+\left.\frac{1}{2}[({\bf g}_1\cdot{\bf g}^\prime)({\bf g}_2-{\bf g}^{\prime\prime})-
({\bf g}_2\cdot{\bf g}^{\prime\prime})({\bf g}_1-{\bf g}^{\prime}) +
({\bf g}^\prime\cdot{\bf g}^{\prime\prime})({\bf g}_1-{\bf g}_2)    ]  \right \}.~~~~~~~~~~~~
\label{31}
\end{eqnarray}
\end{widetext}
In the process of derivation of these equations it was important to keep all the integrations in the collision integral which allows  us to demonstrate the vanishing of its imaginary part due to symmetry ${\bf k}^\prime \leftrightarrow{\bf k}^{\prime\prime} $.  

Now, integrating over $d^3{\bf k}^\prime$ we liquidate the $\delta$ function of momenta. 
Following usual linearization procedure we substitute in the left hand sides of Eqs.(\ref{Tr}) and (\ref{Trs}) the equilibrium distribution Eq.(\ref{eqv}) and on the right hand side we leave only the terms linear in deviation from equilibrium distribution. Thus, we obtain the  linear  integral equations for the functions 
\begin{eqnarray}
\delta f=f-(n_++n_-),\\\delta{\bf g}={\bf g}-\frac {\mbox{\boldmath$\gamma$}}{ |\mbox{\boldmath$\gamma$}|}(n_+-n_-).
\end{eqnarray} 
To establish the parametrical dependence of relaxation time we don't need, however,   to solve this difficult problem. 
For this purpose it is enough to keep only the terms
with $\delta f_1$ and $\delta{\bf g}_1$ neglecting other terms containing $\delta f^\prime, \delta{\bf g}^\prime$ etc  This is a sort of relaxation time approximation. 
Thus, we come to the following equations
\begin{widetext}
\begin{eqnarray}
e{\bf E}\cdot{\bf v}_1\frac{\partial(n_{1+}+n_{1-})}{\partial \varepsilon_1}=-\frac{\pi}{4}W_0\int \frac{d{\bf k}^{\prime\prime}}{(2\pi)^3}\frac{d{\bf k}_2}{(2\pi)^3}\delta(\varepsilon_{1}+\varepsilon_{2}-\varepsilon^\prime-
\varepsilon^{\prime\prime}) 
\times\nonumber\\
\left \{\left [ f_0^\prime f_0^{\prime\prime}\left (1-\frac{1}{2}f_{02} \right )+\left (1-\frac{1}{2}f_0^\prime \right )\left (1-\frac{1}{2}f_0^{\prime\prime}\right )2f_{02}\right ]\delta f_{1}- {\bf g}_{02}\cdot{\bf g}_0^{\prime\prime}\delta f_1+(f_0^{\prime\prime}-f_{02}){\bf g}_0^\prime\cdot\delta{\bf g}_1 \right.\nonumber\\
\left.- ({\bf g}_0^{\prime\prime}-{\bf g}_{02})\cdot{\bf g}_0^\prime\delta f_1-(2-f_0^\prime-f_0^{\prime\prime}){\bf g}_{02}\cdot\delta{\bf g}_1-(f_0^\prime-f_{02}){\bf g}_0^{\prime\prime}\cdot\delta{\bf g}_1  \right \},
\label{32}
\end{eqnarray}
\begin{eqnarray}
e{\bf E}\cdot{\bf v}_1\frac{\partial(n_{1+}-n_{1-})}{\partial \varepsilon_1}\frac{{\bf k}_1}{|{\bf k}_1|}+
\frac{(n_{1+}-n_{1-})}{|{\bf k}_1|}\left  (e{\bf E}-\frac{(e{\bf E}\cdot{\bf k}_1){\bf k}_1}{{\bf k}_1^2}  \right )
=
-\frac{\pi}{4}W_0\int \frac{d{\bf k}^{\prime\prime}}{(2\pi)^3}\frac{d{\bf k}_2}{(2\pi)^3}\delta(\varepsilon_{1}+\varepsilon_{2}-\varepsilon^\prime-
\varepsilon^{\prime\prime}) \times
\nonumber\\
\left \{\left [ f_0^\prime f_0^{\prime\prime}\left (1-\frac{1}{2}f_{02} \right )+\left (1-\frac{1}{2}f_0^\prime \right )\left (1-\frac{1}{2}f_0^{\prime\prime}\right )2f_{02}\right ]\delta {\bf g}_1- {\bf g}_{02}\cdot{\bf g}_0^{\prime\prime}\delta {\bf g}_1+(f_0^{\prime\prime}-f_{02}){\bf g}_0^\prime\delta f_1 \right.\nonumber\\
\left.-({\bf g}_0^{\prime\prime}-{\bf g}_{02})f^\prime\delta f_{1}-(2{\bf g}_{02}-f_0^{\prime\prime}{\bf g}_{02}-f_2{\bf g}_0^{\prime\prime})\delta f_1-({\bf g}_0^\prime-{\bf g}_{02})\cdot{\bf g}_0{\prime\prime}\delta{\bf g}_1\right \}.
\label{33}
\end{eqnarray}
\end{widetext}
Several terms in Eqs.(\ref{32}) and (\ref{33})  vanishes  at integration over  directions of   momenta. Other subintegrand terms like $\propto(f^{\prime\prime}-f_2){\bf g}^{\prime\prime}\cdot\delta{\bf g}_1$ cancel out due to antisymmetry in respect of interchange of arguments $\varepsilon_2\leftrightarrow\varepsilon^{\prime\prime}$. 
We obtain
\begin{widetext}
\begin{eqnarray}
e{\bf E}\cdot{\bf v}_1\frac{\partial(n_{1+}+n_{1-})}{\partial \varepsilon_1}=-\frac{\pi}{4}W_0\int  \frac{d{\bf k}^{\prime\prime}}{(2\pi)^3}\frac{d{\bf k}_2}{(2\pi)^3}\delta(\varepsilon_{1}+\varepsilon_{2}-\varepsilon^\prime-
\varepsilon^{\prime\prime})\times \nonumber\\
\left \{ f_0^\prime f_0^{\prime\prime}\left (1-\frac{1}{2}f_{02} \right )+\left (1-\frac{1}{2}f_0^\prime \right )\left (1-\frac{1}{2}f_0^{\prime\prime}\right )2f_{02}\right \}\delta f_{1}
\label{34}
\end{eqnarray}
\begin{eqnarray}
e{\bf E}\cdot{\bf v}_1\frac{\partial(n_{1+}-n_{1-})}{\partial \varepsilon_1}\frac{{\bf k}_1}{|{\bf k}_1|}+
\frac{(n_{1+}-n_{1-})}{|{\bf k}_1|}\left  (e{\bf E}-\frac{(e{\bf E}\cdot{\bf k}_1){\bf k}_1}{{\bf k}_1^2}  \right )=
-\frac{\pi}{4}W_0\int 
 \frac{d{\bf k}^{\prime\prime}}{(2\pi)^3}\frac{d{\bf k}_2}{(2\pi)^3}\delta(\varepsilon_{1}+\varepsilon_{2}-\varepsilon^\prime-
\varepsilon^{\prime\prime}) \times\nonumber\\
\left \{ f_0^\prime f_0^{\prime\prime}\left (1-\frac{1}{2}f_{02} \right )+\left (1-\frac{1}{2}f_0^\prime \right )\left (1-\frac{1}{2}f_0^{\prime\prime}\right )2f_{02}
-
{\bf g}_0^\prime\cdot{\bf g}_0{\prime\prime}
\right \}\delta{\bf g}_1~~~~~~~~~~~~~~~~~~~~~
\label{35}
\end{eqnarray}
\end{widetext}
Following the procedure of Ref.3, reproduced in Ref.16 in a somewhat different manner,
 we reexpress the integration over ${\bf k}^{\prime\prime} $ and ${\bf k}_2 $ as 
\begin{equation}
d^3{\bf k}^{\prime\prime}d^3{\bf k}_2\approx\frac{m^{3}}{2\cos(\theta/2)}
d\varepsilon^{\prime\prime}d\varepsilon_{2}d\varepsilon^\prime\sin\theta
d\theta d\phi d\phi_2.
\end{equation}
Here $\theta$ is the angle between ${\bf k}_1$ and ${\bf k}_2 $, $\phi$ is the azimuthal angle of ${\bf k}_2 $ around direction ${\bf k}_1$, and $\phi_2$ is the angle between the planes 
$({\bf k}_1,{\bf k}_2) $ and $({\bf k}^\prime,{\bf k}^{\prime\prime}) $. Due to the Fermi surfaces separation this formula is valid within an accuracy of the terms of the order of $\gamma k_F/\mu$. Taking into account the $\delta$ function of energies we can integrate over $\varepsilon^\prime$ and using the equality
$
\varepsilon^\prime=\varepsilon_1+\varepsilon_2-\varepsilon^{\prime\prime}.
$
we come to the equations
\begin{widetext}
\begin{eqnarray}
e{\bf E}\cdot{\bf v}_1\frac{\partial(n_{1+}+n_{1-})}{\partial \varepsilon_1}=-\frac{\pi}{4}\tilde W_0m^3\int 
d\varepsilon^{\prime\prime}d\varepsilon_{2}
\left \{ f_0^\prime f_0^{\prime\prime}\left (1-\frac{1}{2}f_{02} \right )+\left (1-\frac{1}{2}f_0^\prime \right )\left (1-\frac{1}{2}f_0^{\prime\prime}\right )2f_{02}\right \}\delta f_{1},
\label{44}
\end{eqnarray}
\begin{eqnarray}
e{\bf E}\cdot{\bf v}_1\frac{\partial(n_{1+}-n_{1-})}{\partial \varepsilon_1}\frac{{\bf k}_1}{|{\bf k}_1|}+
\frac{(n_{1+}-n_{1-})}{|{\bf k}_1|}\left  (e{\bf E}-\frac{(e{\bf E}\cdot{\bf k}_1){\bf k}_1}{{\bf k}_1^2}  \right )=~~~~~~~~~~~~~~~~~~~~~~\nonumber\\
-\frac{\pi}{4}\tilde W_0m^3\int 
d\varepsilon^{\prime\prime}d\varepsilon_{2}
\left \{ f_0^\prime f_0^{\prime\prime}\left (1-\frac{1}{2}f_{02} \right )+\left (1-\frac{1}{2}f_0^\prime \right )\left (1-\frac{1}{2}f_0^{\prime\prime}\right )2f_{02}\right \}\delta{\bf g}_1\nonumber\\
+\frac{\pi}{4}W_0m^3\int 
d\varepsilon^{\prime\prime}d\varepsilon_{2}\int\frac{\cos\frac{\theta}{2}d\theta d\phi d\phi_2}{(2\pi)^6}
({\bf g}_0^\prime\cdot{\bf g}_0{\prime\prime})
\delta{\bf g}_1,~~~~~~~~~~~~~~~~~~~~~
\label{45}
\end{eqnarray}
where $\tilde W_0=W_0\int\cos\frac{\theta}{2}d\theta d\phi d\phi_2/(2\pi)^6$. These equations contain the integral
\begin{eqnarray}
I_1(\varepsilon_1)=\int d\varepsilon^{\prime\prime}d\varepsilon_{2}
\left \{ f_0^\prime f_0^{\prime\prime}\left (1-\frac{1}{2}f_{02} \right )+\left (1-\frac{1}{2}f_0^\prime \right )\left (1-\frac{1}{2}f_0^{\prime\prime}\right )2f_{02}\right \}
\nonumber\\
=\frac{1}{2}\sum_{\lambda,\mu,\nu}\int  d\varepsilon^{\prime\prime}d\varepsilon_{2}\left \{ n^\prime_\lambda n^{\prime\prime}_\mu(1-n_{2\nu})+ 
(1-n^\prime_\lambda)(1- n^{\prime\prime}_\mu)n_{2\nu}\right \},
\end{eqnarray}
\end{widetext}
where, the indices $\lambda,\mu,\nu=\pm$.

Let us calculate one particular term in this sum
\begin{eqnarray}
\frac{1}{2}
\int d\varepsilon^{\prime\prime}d\varepsilon_2 n^\prime_+ n^{\prime\prime}_+(1-n_{2+})=~~~~~~~~~~~~~~~\nonumber\\
\frac{1}{2}T^2\int dxdy
\frac{1}{e^{t_++x-y}+1}\frac{1}{e^y+1}\frac{1}{e^{-x}+1}=~~~~~~~~~\nonumber\\
\frac{1}{2}T^2\int dx\frac{t_++x}{e^{t_++x}+1}\frac{1}{e^{-x}+1}=\frac{1}{2}T^2\frac{\pi^2+t^2_+}{2}n(t_+).~~~
\end{eqnarray}
Here 
\begin{equation}
x=\frac{\varepsilon_2-\mu_+}{T},~~~y=\frac{\varepsilon^{\prime\prime}-\mu_+}{T},~~~~t_+=\frac{\varepsilon_1-\mu_+}{T}, 
\end{equation}
and
\begin{equation}
n(t_+)=\frac{1}
{e^{t_+}+1} 
\end{equation}
is the Fermi distribution function. 
Similar integration in all the other terms yields
\begin{widetext}
\begin{eqnarray}
I_1(\varepsilon_1)=\frac{1}{2}\sum_{\lambda,\mu,\nu} \int d\varepsilon^{\prime\prime}d\varepsilon_{2}\left \{ n^\prime_\lambda n^{\prime\prime}_\mu(1-n_{2\nu})
 + (1-n^\prime_\lambda0(1- n^{\prime\prime}_\mu)n_{2\nu}\right \}=~~~~~~~~~~~~~~~~~~~~~~~~~\nonumber\\
\frac{1}{4}T^2\left\{[\pi^2+t_+^2]n(t_+)+[\pi^2+(t_+-\kappa)^2]n(t_+-\kappa)
+[\pi^2+(t_++\kappa)^2]n(t_++\kappa)+[\pi^2+t_-^2]n(t_-)\right.\nonumber\\
\left.+[\pi^2+t_+^2]n(t_+)+
[\pi^2+(t_--\kappa)^2]n(t_--\kappa)
+[\pi^2+(t_-+\kappa)^2]n(t_-+\kappa)
+[\pi^2+t_-^2]n(t_-)\right\} \nonumber\\
+\left (~~t_\pm\to-t_\pm,~~\kappa\to-\kappa~~
\right )~~~~~~~~~~~~~~~~~~~~~~~~~~~~~~~~~~~~~~~~~~~~~~~~~~~~~\nonumber\\
=\frac{1}{4}T^2\left\{[\pi^2+t_+^2]+[\pi^2+(t_+-\kappa)^2]
+[\pi^2+(t_++\kappa)^2]+[\pi^2+t_-^2]~~~~~~~~~~~~\right.\nonumber\\
\left.+[\pi^2+t_+^2]+
[\pi^2+(t_--\kappa)^2]
+[\pi^2+(t_-+\kappa)^2]
+[\pi^2+t_-^2]\right\},~~~~~~~~~~~~~~~~~~
\end{eqnarray}
\end{widetext}
where we used the notation $\kappa=(\mu_+-\mu_-)/T$.
Finally we obtain
\begin{equation}
I_1(\varepsilon_1)=2(\pi T)^2+(\varepsilon_1-\mu_+)^2 +(\varepsilon_1-\mu_-)^2+(\mu_+-\mu_-)^2.
\label{I1}
\end{equation}
Thus, the deviation of the scalar part of distribution function from its equilibrium value is
\begin{equation}
\delta f({\bf k}_1)=-\frac{4e}{\pi \tilde W_0m^3
I_1(\varepsilon_1)} {\bf E}\cdot{\bf v}_1
\frac{\partial(n_{1+}+n_{1-})}{\partial \varepsilon_1}
\label{scalar}
\end{equation}

The second integral  in Eq.(\ref{45}) is 
\begin{widetext}
\begin{eqnarray}
I_2=\int 
d\varepsilon^{\prime\prime}d\varepsilon_{2}\int\frac{\cos\frac{\theta}{2}d\theta d\phi d\phi_2}{(2\pi)^6}
({\bf g}_0^\prime\cdot{\bf g}_0{\prime\prime})
= ~~~~~~~~~~~~~~~~~~~~~~~~~~~~~~~~\nonumber\\
\int d\varepsilon^{\prime\prime}d\varepsilon_{2}\int\frac{\cos\frac{\theta}{2}d\theta d\phi d\phi_2}{(2\pi)^6}
\frac{{\bf k}_1+{\bf k}_2-{\bf k}^{\prime\prime}-2\pi{\bf m}/a}{|{\bf k}_1+{\bf k}_2-{\bf k}^{\prime\prime}-2\pi{\bf m}/a|}
\cdot\frac{{\bf k}^{\prime\prime}}
{|{\bf k}^{\prime\prime}|}(n_{0+}^\prime-n_{0-}^\prime)(n_{0+}^{\prime\prime}-n_{0-}^{\prime\prime}).
\end{eqnarray}
\end{widetext}
The analytic calculation of this integral is impossible. However, at $\mu^+\to\mu_-$ it tends to zero as
$\propto (\mu^+-\mu_-)^2$. We will treat this integral as energy and momentum independent  constant
\begin{equation}
I_2\approx (\mu^+-\mu_-)^2.
\label{I2}
\end{equation}

Thus, the deviation of the spinor part of the distribution function from its equilibrium value is
\begin{widetext}
\begin{equation}
\delta{\bf g}({\bf k}_1)=-\frac{4e}{\pi m^3\left [\tilde W_0
I_1(\varepsilon_1)-W_0I_2\right ]}\left [{\bf E}\cdot{\bf v}_1\frac{\partial(n_{1+}-n_{1-})}{\partial \varepsilon_1}\hat{\bf k}_1+
\frac{(n_{1+}-n_{1-)}}{|{\bf k}_1|}\left  ({\bf E}-({\bf E}\cdot\hat{\bf k}_1)\hat{\bf k}_1  \right )\right ],
\label{spinor}
\end{equation}\end{widetext}
where $\hat{\bf k}_1=
\frac{{\bf k}_1}{|{\bf k}_1|}$.


\end{document}